\documentclass[12pt]{article}
\pdfoutput=1

\usepackage{amsmath,amssymb,amsfonts,graphicx,amsfonts}
\usepackage[hidelinks]{hyperref}
\usepackage{ascmac}
\usepackage{physics}

\allowdisplaybreaks[4]

\date{}
\begin{document}
\begin{flushright}
\today\\
\end{flushright}

\vspace{0.1cm}

\begin{center}
  {\Large
Estimating truncation effects of quantum bosonic systems\\
using sampling algorithms
}
\end{center}
\vspace{0.1cm}
\vspace{0.1cm}
\begin{center}

Masanori Hanada$^{a,*}$, Junyu Liu$^b$, Enrico Rinaldi$^c$, Masaki Tezuka$^d$ \footnote{Authors in alphabetical order.}

\vspace{0.3cm}

{\small
$^a$School of Mathematical Sciences, Queen Mary University of London,
Mile End Road, London, E1 4NS, UK\\
$^{a,b}$qBraid Co., Harper Court 5235, Chicago, IL 60615, USA\\
$^b$Pritzker School of Molecular Engineering, The University of Chicago, Chicago, IL 60637, USA\\
$^b$Chicago Quantum Exchange, Chicago, IL 60637, USA\\
$^b$Kadanoff Center for Theoretical Physics, The University of Chicago, Chicago, IL 60637, USA\\
$^c$
Quantinuum K.K., Otemachi Financial City Grand Cube 3F, 1-9-2 Otemachi,\\
Chiyoda-ku, Tokyo, Japan\\
$^c$Physics  Department,  University  of  Michigan,
Ann  Arbor,  MI  48109, USA\\
$^c$Theoretical  Quantum  Physics  Laboratory, RIKEN,
Wako, Saitama 351-0198, Japan\\
$^c$Interdisciplinary Theoretical and Mathematical Sciences (iTHEMS) Program, RIKEN,\\
Wako, Saitama 351-0198, Japan\\
$^d$Department of Physics, Kyoto University, Kyoto 606-8502, Japan
}\\
* Author to whom any correspondence should be addressed.\\
\textbf{E-mail:}
m.hanada@qmul.ac.uk, 
junyuliucaltech@gmail.com,
erinaldi.work@gmail.com,
and 
tezuka@scphys.kyoto-u.ac.jp.
\end{center}

\vspace{1.5cm}

\begin{center}
  {\bf Abstract}
\end{center}
To simulate bosons on a qubit- or qudit-based quantum computer, one has to regularize the theory by truncating infinite-dimensional local Hilbert spaces to finite dimensions.
In the search for practical quantum applications, it is important to know how big the truncation errors can be.
In general, it is not easy to estimate errors unless we have a good quantum computer.
In this paper, we show that traditional sampling methods on classical devices, specifically Markov Chain Monte Carlo, can address this issue for a rather generic class of bosonic systems with a reasonable amount of computational resources available today.
As a demonstration, we apply this idea to the scalar field theory on a two-dimensional lattice, with a size that goes beyond what is achievable using exact diagonalization methods.
This method can be used to estimate the resources needed for realistic quantum simulations of bosonic theories, and also, to check the validity of the results of the corresponding quantum simulations.

\newpage
\tableofcontents

\section{Introduction}\label{sec:introduction}
\hspace{0.51cm}
The Hilbert space for bosons is infinite-dimensional.
To simulate bosons on a qubit- or qudit-based quantum computer, one has to introduce a finite-dimensional approximation of the theory by truncating the Hilbert space~\cite{Jordan:2011ci,Jordan:2012xnu,Macridin:2018oli,Macridin:2018gdw,Klco:2018zqz,Macridin:2021uwn,Alexandru:2022son,Li:2022ped}.
Sometimes this truncation is referred to as digitization and it is one of the necessary steps in the construction of efficient quantum algorithms to simulate quantum gauge theories~\cite{Lamm:2019bik,Ciavarella:2021nmj,Alexandru:2019nsa,Ji:2020kjk,Ji:2022qvr}, and to estimate quantum resources~\cite{Hackett:2018cel,Mathis:2020fuo,Alexandru:2021jpm}.
We will use truncation and digitization to mean the same thing in the rest of the manuscript, and this should not be confused with the discretization of the spacetime continuum on the lattice.

Some errors associated with digitization are inevitable, and one needs to know how big they can be.
While previous studies on truncated bosonic Hilbert spaces~\cite{Jordan:2011ci,Jordan:2012xnu,Macridin:2018oli,Macridin:2018gdw,Klco:2018zqz,Macridin:2021uwn,Alexandru:2022son,Li:2022ped} have looked at the scaling of the errors with the digitization spacing (or equivalently, the number of qubits), this does not seem to be a straightforward task in general.
In fact, when the bosonic systems investigated are too large, one would need to run the digitized theory on a good (noiseless) quantum computer and also efficiently compute the same observable with classical algorithms in order to benchmark the result.
So far, quantitative assessment of the digitization errors using classical devices is limited to rather small systems because the only known approaches applicable to generic theories are based on the explicit construction of Hilbert space whose dimensionality increases exponentially with respect to the system size.
A thorough numerical study of digitization errors for a $\phi^4$ lattice scalar theory was done in Ref.~\cite{Macridin:2021uwn}, where the authors studied a single-site (one bosonic degree of freedom) and a two-site (two bosonic degrees of freedom) system.
However, it was difficult to go to larger lattice sizes, due to the exponential growth of memory resources with the number of lattice sites (bosonic degrees of freedom).
In this paper, we will resolve this issue for a rather generic class of bosonic theories and reduce the required resources from exponential to polynomial when estimating expectation values.
We point out that this is the only known method able to do so for a rather generic class of bosonic systems in arbitrary dimensions.
As a modest demonstration, we will study the digitization effect on a two-dimensional 16-site lattice model (16 bosonic degrees of freedom).

It is widely believed that the truncation effect vanishes exponentially, e.g., the low-lying energy eigenvalues have correction suppressed exponentially with respect to the truncation level~\cite{somma2015quantum,Macridin:2018gdw,Macridin:2018oli,Klco:2018zqz,Rinaldi:2021jbg,Liu:2021ohs,Bauer:2022hpo,Liu:2021otn}.
Although this scaling has been verified for concrete examples, there is no proof applicable to a wide class of theories.
Furthermore, there is an implicit assumption, i.e., the wave function decays exponentially fast as $|x|$ becomes large.
Note also that, even if the exponential accuracy of the digitization is valid, the precise value of the error for each theory is not immediately clear.
However, it is practically impossible to study the digitization effects via the explicit construction of the Hilbert space except for small systems, e.g. lattice field theories with a few local degrees of freedom and less than a dozen lattice sites~\cite{Klco:2018zqz,Macridin:2021uwn}.
It is a frustrating situation that prevents us from quantitative resource estimates for quantum simulations. 
A related issue is that it is not easy to check the validity of the results of the quantum simulation with specific digitization on NISQ devices: the presence of noise can quickly invalidate the results without a working error-correction strategy.
It would be practically useful if we can do some calculations on classical devices to establish robust truncation error quantification which, as a byproduct, offers us some ways to cross-check quantum simulations on NISQ devices.\footnote{
Note that the MCMC method can be used only for a limited class of quantities. Although such quantities enable us to cross-check the validity of the simulation (e.g., we can check if the correct ground-state wave function is obtained), the MCMC method cannot replace the quantum simulation (e.g., it is impossible to add a small excitation to the ground state and determine the Hamiltonian time evolution).
}

In this paper, we show that a Markov Chain Monte Carlo (MCMC) method can be used to estimate the digitization effects in a rather generic class of bosonic systems.
MCMC methods~\cite{Metropolis:1953am} (see Ref.~\cite{Hanada-Matsuura} for an elementary introduction) are straightforward in the original bosonic theory before digitization. 
Indeed, one can use the Euclidean path-integral method combined with MCMC to study some features of the quantum systems as long as there is no sign problem. 
If similar simulations are doable for the digitized theory, one can estimate the digitization effects. 

A caveat is that the absence of the sign problem in the Euclidean path integral without digitization does not necessarily guarantee the absence of the sign problem in the digitized theory.
In this work, we will show that we can apply MCMC to a wide class of theories without having the sign problem, at least for a certain digitization scheme.
As a result, we are able to properly assess the amount of digitization effects even for large systems, well beyond the limits of exact diagonalization techniques.

We consider the generic system of $N_{\rm bos}$ bosons consisting of coordinate variables $\hat{x}_i$ ($i=1,2,\cdots,N_{\rm bos}$) and conjugate momentum variables $\hat{p}_i$ that satisfy the canonical commutation relation\footnote{In our convention, $\hbar=1$.} 
\begin{align}\label{eq:commutation-relation}
    [\hat{x}_j,\hat{p}_k]=i\delta_{jk}\ . 
\end{align}

We assume that the Hamiltonian is given by  
\begin{align}\label{eq:general-bosonic-Hamiltonian}
    \hat{H}=\frac{1}{2}\sum_{i=1}^{N_{\rm bos}}\hat{p}_i^2+V(\hat{x}_1,\cdots,\hat{x}_{N_{\rm bos}})\ ,  
\end{align}
where $V(x_1,\cdots,x_{N_{\rm bos}})$ is a real function bounded from below and represents the potential energy as a function of the coordinate variables only.
This is a rather generic class and it even includes SU($N$) Yang-Mills theories described by using the orbifold lattice formulation~\cite{Kaplan:2002wv,Buser:2020cvn}. 
We use the truncation in the coordinate basis defined in Sec.~\ref{sec:digitization}. 
As we will see, this scheme admits the MCMC simulations without sign problem. 

This paper is organized as follows.
In Sec.~\ref{sec:digitization}, we introduce the digitization scheme associated with the coordinate basis. 
Firstly, the case of a single bosonic variable is explained, and then it is generalized to the case of a generic number of variables.
The digitization of $(2+1)$-dimensional scalar Quantum Field Theory (QFT) on a lattice is explained as a concrete example. 
In Sec.~\ref{sec:mcmc}, the Monte Carlo technique is introduced and some numerical experiments are conducted. 
Sec.~\ref{sec:conclusion} is devoted to concluding remarks and discussion of future directions. 

\section{Truncation scheme}\label{sec:digitization}
\hspace{0.51cm}
In this section, we specify the truncation scheme. 
\subsection{Coordinate-basis truncation for the single-boson system}
\hspace{0.51cm}

Here, we introduce digitization in the coordinate basis~\cite{Jordan:2012xnu,Jordan:2011ci,Klco:2018zqz}, also called the field amplitude basis~\cite{Macridin:2021uwn,Li:2022ped}. 
Let us start with a review of the single boson field example. 
Let $\{\ket{x}|x\in\mathbb{R}\}$ be the coordinate basis for this particle that satisfies
\begin{align}\label{eq:coordinate-operator}
    \hat{x}\ket{x} = x\ket{x} \, . 
\end{align}
A simple way to digitize it is to introduce a cutoff to the value of the eigenvalue $x$ as 
\begin{align}\label{eq:ir-cutoff}
    -R\le x\le R \, , 
\end{align}
and introduce $\Lambda$ points, 
\begin{align}\label{eq:coordinate-spacing}
    x(n) = -R+na_{\rm dig} \, ,
    \qquad
    a_{\rm dig} = \frac{2R}{\Lambda-1} \, , 
    \qquad
    n=0,1,\cdots,\Lambda-1 \, .   
\end{align}

The digitization parameters $\Lambda$, $a_{\rm dig}$, and $R$ should be sent to infinity, zero, and infinity, respectively, to recover the action of the original operator $\hat{x}$.
By using $\ket{n}$ to denote $\ket{x(n)}$, we can write
\begin{align}\label{eq:digitized-coordinate-operator}
    \hat{x} = \sum_{n=0}^{\Lambda-1} x(n) \ket{n}\bra{n}. 
\end{align}

The momentum operator $\hat{p}$ appears in the Hamiltonian only in the form of $\hat{p}^2$.
A convenient way of regularizing it is 
\begin{align}\label{eq:digitized-momentum-square-operator}
    \hat{p}^2 =
    \frac{1}{a_{\rm dig}^2}
    \left\{
        \sum_{n=0}^{\Lambda-1}
        2\ket{n}\bra{n}
        -
        \sum_{n=0}^{\Lambda-2}
        \ket{n+1}\bra{n}
        -
        \sum_{n=0}^{\Lambda-2}
        \ket{n}\bra{n+1}
    \right\} \, . 
\end{align}

The dimension of the Hilbert space is $\Lambda$ and the truncated Hamiltonian is expressed as a $\Lambda\times\Lambda$ matrix.
For each concrete example, we can diagonalize the Hamiltonian up to a rather large value of $\Lambda$ and confirm that the digitization effects below a fixed energy scale disappear exponentially as $\sim e^{-c\Lambda}$ with some $c>0$.
In terms of the number of qubits $q$, the truncation level is $\Lambda=2^q$, and hence the suppression of the digitization effects is expected to be doubly exponential, $\sim e^{-c\cdot 2^q}$. 

\subsection{Coordinate-basis truncation for multi-boson system}
\hspace{0.51cm}
Next, let us consider the more interesting case where we have multiple bosonic degrees of freedom.
Suppose there are $N_{\rm bos}$ variables $\vec{x}=(x_1,\cdots,x_{N_{\rm bos}})\in\mathbb{R}^{N_{\rm bos}}$. 
We introduce $N_{\rm bos}$ integers $\vec{n}=(n_1,\cdots,n_{N_{\rm bos}})\in\{0,1,\cdots,\Lambda-1\}^{N_{\rm bos}}$ that are related to $x_i(n_i)$ as 
\begin{align}\label{eq:digitized-positions}
    x_i(n_i) = -R + n_i a_{\rm dig} \, ,
    \qquad
    a_{\rm dig}=\frac{2R}{\Lambda-1} \, , 
    \qquad
    n_i=0,1,\cdots,\Lambda-1 \, .   
\end{align}

Note that we can take different $R$, $\Lambda$, and $a_{\rm dig}$ for each bosonic coordinate $x_i$; here we use the same values for simplicity.
For each $i=1,2,\cdots,N_{\rm bos}$, the momentum operator $\hat{p}_i$ appears in the Hamiltonian only in the form of $\hat{p}_i^2$.
This is a natural extension of the single-particle case presented in the previous section.
Again we choose a regularization 
\begin{align}\label{eq:digitized-momentum-square-operator-bosons}
    \hat{p}_i^2
    =
    \frac{1}{a_{\rm dig}^2}
    \left\{
        \sum_{n=0}^{\Lambda-1}
        2\ket{\vec{n}}\bra{\vec{n}}
        -
        \sum_{n=0}^{\Lambda-2}
        \ket{\vec{n}+\hat{i}}\bra{\vec{n}}
        -
        \sum_{n=0}^{\Lambda-2}
        \ket{\vec{n}}\bra{\vec{n}+\hat{i}}
    \right\} \, . 
\end{align}

The dimension of the Hilbert space is $\Lambda^{N_{\rm bos}}$ and the truncated Hamiltonian is expressed as a $\Lambda^{N_{\rm bos}}\times\Lambda^{N_{\rm bos}}$ matrix. 
Unless $N_{\rm bos}$ is relatively small, it is difficult to directly determine the energy eigenvalues.
Still, it is believed that the digitization effects below a fixed energy scale disappear exponentially as $\sim e^{-c\Lambda}$ with some $c>0$~\cite{Klco:2018zqz,Macridin:2021uwn}.
As far as we know, there is no proof for this scaling.
(Here, an implicit assumption is that the wave function decays exponentially fast as $|x|$ becomes large.) 
Note also that, even if the exponential accuracy of the digitization is valid, the precise dependence of the errors on $a_{\rm dig}$ for each theory is not immediately clear.
Therefore, it is important to have a method to make a quantitative estimation of the truncation effect that is applicable to a wide class of theories.
Our solution to this problem in presented later in Sec.~\ref{sec:mcmc}.

\subsection{Scalar QFT on spatial lattice}
\hspace{0.51cm}
As a particularly important example and prototypical QFT, we consider a scalar QFT on a $d$-dimensional spatial lattice.
We consider the square lattice with equal lattice spacing $a_{\rm lat}$ in all $d$ directions.
The lattice Hamiltonian is defined by
\begin{align}\label{eq:2d-lattice-Hamiltonian}
    \hat{H}_{\rm lat}
    =
    a_{\rm lat}\hat{H}
    =
    \sum_{\vec{n}_{\rm lat}}\left(
    \frac{1}{2}
    \hat{\pi}_{\vec{n}_{\rm lat}}^2
    +
    \frac{1}{2}
    \sum_{\mu=1}^d
    \left(
    \hat{\phi}_{\vec{n}_{\rm lat}+\hat{\mu}}-\hat{\phi}_{\vec{n}_{\rm lat}}
    \right)^2
    +
    V(\hat{\phi}_{\vec{n}_{\rm lat}})
    \right) \, . 
\end{align}
Fields $\hat{\phi}$ and $\hat{\pi}$ are dimensionless, and they correspond to the fields in the continuum theory by $\hat{\phi}=a_{\rm lat}^{(d-1)/2}\hat{\phi}_{\rm cont.}$ and $\hat{\pi}=a_{\rm lat}^{(d+1)/2}\hat{\pi}_{\rm cont.}$. Parameters such as mass are also made dimensionless, e.g., $m_{\rm lat}=a_{\rm lat}\times m$. 
A vector $\vec{n}_{\rm lat}\in\mathbb{Z}^d$ labels the lattice sites. 
This is different from $\vec{n}$ used for the digitization, which we now denote by $\vec{n}_{\rm dig}$. 
The canonical commutation relation is imposed, i.e., 
\begin{align}\label{eq:commutation-relation-lattice}
    [\hat{\phi}_{\vec{n}_{\rm lat}},\hat{\pi}_{\vec{n}'_{\rm lat}}]
    =
    i\delta_{\vec{n}_{\rm lat},\vec{n}'_{\rm lat}} \, .
\end{align}

The operators $\hat{\phi}$ and $\hat{\pi}$ are the same as $\hat{x}$ and $\hat{p}$ in the previous sections: here we have used a different notation to more easily connect with the traditional symbols used in the physics community.

$\hat{\mu}$ is the unit vector along the $\mu$-th dimension of the spatial lattice ($\mu=1,\cdots,d$).
This is different from the unit vector $\hat{i}$ used for the digitization ($i=1,\cdots,N_{\rm bos}$).
As infrared regularization, we introduce the periodic boundary condition with period $L$ to all directions. 
Typically, the continuum limit ($a_{\rm lat}\to 0$) is taken by fixing the physical volume $a_{\rm lat}L$. 
The number of bosonic degrees of freedom is $N_{\rm bos}=L^d$.
Hence the dimension of the truncated Hilbert space is $\Lambda^{N_{\rm bos}}=\Lambda^{L^d}$. 

We digitize $\hat{\phi}$ just as before, by introducing $R$, $\Lambda$ and $a_{\rm dig}$. 
We use the same digitization parameters for all lattice points for sake of simplicity. 
Note that $a_{\rm lat}\to 0$ and $a_{\rm dig}\to 0$ do not necessarily commute: the correct order is $a_{\rm dig}\to 0$ first, then $a_{\rm lat}\to 0$.

\section{Monte Carlo estimate for truncation effect}\label{sec:mcmc}
\hspace{0.51cm}
In this section, we show that the MCMC methods~\cite{Metropolis:1953am} (see Ref.~\cite{Hanada-Matsuura} for an elementary introduction) can be used to estimate the digitization effects. 
Numerical demonstrations will be provided as well. 
A short review of the MCMC methods is provided in Appendix~\ref{sec:MCMC}. 
The crucial point is that, as long as the problem under consideration reduces to an average over non-negative weights, the MCMC methods allow efficient computations. 

\subsection{Formulation and algorithm}\label{sec:MCMC-formulation}
\hspace{0.51cm}
Let us estimate the truncation effect in the coordinate-basis scheme by using Markov Chain Monte Carlo simulation. 
For $N_{\rm bos}$ variables $\vec{x}=(x_1,\cdots,x_{N_{\rm bos}})$, we define integers $\vec{n}=(n_1,\cdots,n_{N_{\rm bos}})$ via the relation \eqref{eq:digitized-positions}. 
For a Hamiltonian defined by \eqref{eq:general-bosonic-Hamiltonian}, we consider the thermal partition function defined by 
\begin{align}\label{eq:partition-function}
    Z(\beta) = {\rm Tr} \, e^{-\beta\hat{H}} \, .
\end{align}

Here, $\beta$ is related to the temperature $T$ by $\beta=\frac{1}{T}$.
The trace is over the truncated Hilbert space. 
We rewrite it as
\begin{align}\label{eq:partition-function-Trotter}
    Z(\beta) & =
    \sum_{\vec{n}^{(1)},\cdots,\vec{n}^{(K)}}
        \bra{\vec{n}^{(1)}}e^{-\Delta\cdot\hat{H}} \ket{\vec{n}^{(2)}}
    \cdot
        \bra{\vec{n}^{(2)}}e^{-\Delta\cdot\hat{H}} \ket{\vec{n}^{(3)}}
    \cdots
        \bra{\vec{n}^{(K)}}e^{-\Delta\cdot\hat{H}} \ket{\vec{n}^{(1)}}
    \, ,
\end{align}
where $\beta=\Delta \times K$ has been divided up into $K$ intervals.
If $\Delta$ is sufficiently small, we can rewrite $\bra{\vec{n}^{(j)}}e^{-\Delta\cdot\hat{H}} \ket{\vec{n}^{(j+1)}}$ as
\begin{align}\label{eq:Trotter-intermediate}
    \bra{\vec{n}^{(j)}}e^{-\Delta\cdot\hat{H}} \ket{\vec{n}^{(j+1)}}
    &\simeq
    \bra{\vec{n}^{(j)}}e^{-\Delta\cdot\sum_i\frac{\hat{p}_i^2}{2}}e^{-\Delta\cdot V(\hat{x})} \ket{\vec{n}^{(j+1)}}
    \nonumber\\
    &=
    \bra{\vec{n}^{(j)}}e^{-\Delta\cdot\sum_i\frac{\hat{p}_i^2}{2}} \ket{\vec{n}^{(j+1)}}
    \cdot e^{-\Delta\cdot V(\vec{n}^{(j+1)})} \, . 
\end{align}

To handle $\bra{\vec{n}^{(j)}}e^{-\Delta\cdot\sum_i\frac{\hat{p}_i^2}{2}} \ket{\vec{n}^{(j+1)}}$, we can truncate the Fourier expansion of $e^{-\Delta\cdot\sum_i\frac{\hat{p}_i^2}{2}}$.

If we keep order-$\Delta$ terms, we obtain
\begin{align}\label{eq:Trotter-full}
    &\simeq
    \bra{\vec{n}^{(j)}}\left(1-\Delta\cdot\sum_i\frac{\hat{p}_i^2}{2}\right) \ket{\vec{n}^{(j+1)}}
    \cdot e^{-\Delta\cdot V(\vec{n}^{(j+1)})}
    \nonumber\\
    &=
    \left\{
    \left(
    1-\frac{N_{\rm bos}\Delta}{a_{\rm dig}^2}
    \right)\delta_{\vec{n}^{(j)},\vec{n}^{(j+1)}}
    +
    \frac{\Delta}{2a_{\rm dig}^2}
    \sum_{i=1}^{N_{\rm bos}}
    \left(
    \delta_{\vec{n}^{(j)},\vec{n}^{(j+1)}+\hat{i}}
    +
    \delta_{\vec{n}^{(j)},\vec{n}^{(j+1)}-\hat{i}}
    \right)
    \right\}
    \cdot e^{-\Delta\cdot V(\vec{n}^{(j+1)})} \, .
\end{align}

The crucial point that enables us to use MCMC methods is that the final form in \eqref{eq:Trotter-full} is non-negative for any $\vec{n}^{(j)}$ and $\vec{n}^{(j+1)}$, if $1-\frac{N_{\rm bos}\Delta}{a_{\rm dig}^2}>0$.
In other words, we can write the partition function as a sum of non-negative weights.

Of course, when the digitization is removed and $x_i$ are treated as continuous variables, we can easily rewrite the partition function in the form of the standard Feynman path integral, which is well known to be amenable to MCMC simulations. 
In some sense, we are solving a simple problem in a complicated manner, to estimate the digitization artifact.
On the other hand, we are using a mature classical numerical simulation technique to understand an issue arising in the new growing field of quantum simulations for quantum field theories. 

Note that $\frac{\Delta}{a_{\rm dig}^2}$ has to be small for \eqref{eq:Trotter-full} to be a good approximation.
Therefore, as $a_{\rm dig}$ becomes smaller, we need smaller $\Delta$ and hence more Trotterization steps. 
For this reason, the classical simulation cost is sensitive to $a_{\rm dig}$, but not to $\Lambda$.
In the demonstrations shown in the later sections, we take $\Lambda$ and $R$ very large so that the finite-$R$ effect is negligible and we can focus on the finite-$a_{\rm dig}$ effect. 

Via the Monte Carlo simulation, expectation values of functions of $\vec{x}$ can be estimated from stochastic samples. 
We write the expectation value as
\begin{align}\label{eq:expectation-value}
    \left\langle f(\vec{x}) \right\rangle_\beta
    =
    \frac{1}{Z(\beta)}
    \sum_{\vec{n}^{(1)},\cdots,\vec{n}^{(K)}}
    \bra{\vec{n}^{(1)}}e^{-\Delta\cdot\hat{H}} \ket{\vec{n}^{(2)}}
    \cdot
    \bra{\vec{n}^{(2)}}e^{-\Delta\cdot\hat{H}} \ket{\vec{n}^{(3)}}
    \cdots
    \bra{\vec{n}^{(K)}}e^{-\Delta\cdot\hat{H}} \ket{\vec{n}^{(1)}}f(\vec{x}) \, . 
\end{align}

Here, $\vec{x}$ and $\vec{n}$ are related by \eqref{eq:digitized-positions}. 
We use the approximation \eqref{eq:Trotter-full} for the simulation.
In the limit of $\Delta\to 0$, the approximation becomes exact, and expectation values at finite temperature can be obtained including the digitization effects.
By dialing the temperature, the digitization effects at various energy scales can be studied. 

\subsubsection*{Metropolis algorithm}
\hspace{0.51cm}
In the Metropolis algorithm, the chain of configurations $\{\vec{n}^{(1)},\cdots,\vec{n}^{(K)}\}$ is generated in such a way that the probability distribution of configurations is proportional to the right-hand side of \eqref{eq:partition-function-Trotter}, where the approximation \eqref{eq:Trotter-full} is understood. 
More explicitly, the probability distribution is proportional to $e^{-S(\vec{n}^{(1)},\cdots,\vec{n}^{(K)};\Delta)}$, where
\begin{align}
    &
    \exp\left(-S(\vec{n}^{(1)},\cdots,\vec{n}^{(K)};\Delta)\right)
    \nonumber\\
    &\equiv
    \prod_{j=1}^K
    \left\{
    \left(
    1-\frac{N_{\rm bos}\Delta}{a_{\rm dig}^2}
    \right)\delta_{\vec{n}^{(j)},\vec{n}^{(j+1)}}
    +
    \frac{\Delta}{2a_{\rm dig}^2}
    \sum_{i=1}^{N_{\rm bos}}
    \left(
    \delta_{\vec{n}^{(j)},\vec{n}^{(j+1)}+\hat{i}}
    +
    \delta_{\vec{n}^{(j)},\vec{n}^{(j+1)}-\hat{i}}
    \right)
    \right\}
    \cdot e^{-\Delta\cdot V(\vec{n}^{(j+1)})} \, . 
\end{align}

Note that 
\begin{align}
    \lim_{K\to\infty}
    \sum_{\vec{n}^{(1)},\cdots,\vec{n}^{(K)}}
    \exp\left(-S(\vec{n}^{(1)},\cdots,\vec{n}^{(K)};\Delta)\right)
    \to Z(\beta)
    \qquad(\beta=K\Delta:{\rm fixed}). 
\end{align}

In the following, we describe a naive update move for the Metropolis algorithm.
The algorithm will start from an initial point, or configuration of all variables.
For the initial configuration, we take all $\vec{n}^{(j)}$ to be the same and between $0$ and $\Lambda-1$.
(Practically, we should take $n\sim\frac{\Lambda}{2}$, $x\sim 0$.)
We perform the following procedure to $j=1,2,\cdots,K=\frac{\beta}{\Delta}$ and $i=1,2,\cdots,N_{\rm bos}$:

\begin{enumerate}
\item 
{\bf Proposal}: as a candidate for the new value of $n_i^{(j)}$, $n'\equiv n_i^{(j)}\pm 1$ (equivalently, $\vec{n}'=\vec{n}^{(j)}\pm\hat{i}$) is proposed with probability $\frac{1}{2}$ for each.

\item
The candidate $n'$ is automatically rejected (i.e., $n_i^{(j)}$ remains unchanged) unless the following three conditions are satisfied:
\begin{align}
    & 0\le n'\le \Lambda-1\ ,\\
    & |\vec{n}^{\prime}-\vec{n}^{(j+1)}|=0\ {\rm or}\ 1,\\
    & |\vec{n}^{\prime}-\vec{n}^{(j-1)}|=0\ {\rm or}\ 1.
\end{align}

\item
{\bf Metropolis test}: the candidate $n'$ is accepted (i.e., $n_i^{(j)}$ becomes $n'$) with the probability ${\rm min}\left(1,e^{-\delta S}\right)$, where $\delta S$ is the increment of the action.  
Otherwise the candidate $n'$ is rejected (i.e., $n_i^{(j)}$ remains unchanged).

\end{enumerate}
This is `one sweep'. 
We repeat many sweeps and collect successive configurations along the Markov Chain. 
The above naive approach is not efficient because $\vec{x}$ is changed only locally (i.e., only one time slice at each step), while $\vec{x}$ should slowly change along the imaginary-time circle following dominant configurations. 

In classical MCMC simulations of spin systems, local update rules for the Metropolis algorithm are known to perform poorly, in particular in regions of metastability or close to phase transitions.
Inspired by cluster algorithms~\cite{Swendsen:1987ce}, we improve the algorithm by allowing the simultaneous update of a cluster as follows:
\begin{enumerate}

\item
Choose $1\le B \le B_{\rm max}$ randomly, where $B$ labels the size of the ``clusters'' (or blocks)

\item
{\bf Proposal}: as a candidate for the next configuration, we vary  $n_i^{(j)},n_i^{(j+1)},\cdots,n_i^{(j+B)}$ simultaneously.
For $l=j,j+1,\cdots,j+B$, $n^{\prime (l)}\equiv n_i^{(l)}\pm 1$ (equivalently, $\vec{n}^{\prime (l)}=\vec{n}^{(l)}\pm\hat{i}$) are proposed with probability $\frac{1}{2}$ for each.
Note that we use the same sign for all $l$'s. 

\item
The candidate $n^{\prime}$ is automatically rejected unless the following three conditions are satisfied for all $l=j,\cdots,j+B$:
\begin{align}
    & 0\le n'\le \Lambda-1\ ,\\
    & |\vec{n}^{\prime (j+B)}-\vec{n}^{(j+B+1)}|=0\ {\rm or}\ 1,\\
    & |\vec{n}^{\prime (j)}-\vec{n}^{(j-1)}|=0\ {\rm or}\ 1.
\end{align}

\item
{\bf Metropolis test}: the candidate $n'$ is accepted with the probability ${\rm min}\left(1,e^{-\delta S}\right)$, where $\delta S$ is the increment of the action.  
Otherwise, the candidate $n'$ is rejected. 

\end{enumerate}

$B_{\rm max}$ can be any number between 1 and $K$.
The optimal value depends on the detail of the Hamiltonian and digitization parameters and in principle it can be found via a standard analysis of the autocorrelation time (see e.g., Ref.~\cite{Hanada-Matsuura}).
As an example, one would first try the algorithm with $B_{\rm max}=1$ and measure the autocorrelation time of the observable of interest.
Then, one would increase $B_{\rm max}$ by 1 unit and measure again the autocorrelation time.
For some larger value of $B_{\rm max}$ the autocorrelation time will start decreasing because the update sweeps become more efficient at finding new configurations.
For the simulations reported in this paper, we set $B_{\rm max}=\frac{K}{2}$, except for Fig.~\ref{fig:V-for-quartic-potential_history_non_cluster} in which $B_{\rm max}=1$ is used for comparison.  
We chose $B=1,2,\cdots,B_{\rm max}$ with equal probability. 

\subsubsection*{Simulation cost}
\hspace{0.51cm}

Suppose that $R$ is sufficiently large so that the support of the wave function is mostly contained in the truncated Hilbert space.
As long as this loose condition is satisfied, the simulation cost is not sensitive to $R$.
On the other hand, the cost is sensitive to $a_{\rm dig}$, because it appears in the expansion parameter $\frac{\Delta}{a_{\rm dig}^2}$ in~\eqref{eq:Trotter-full}. 
To keep the expansion parameter small, we need to scale the number of Trotter steps $K$ proportionally to $a_{\rm dig}^{-2}$.
Therefore, the number of variables in the Monte Carlo simulation scales as $N_{\rm bos}K\sim\frac{N_{\rm bos}}{a_{\rm dig}^2}\sim 2^{2n_{\rm q}}N_{\rm bos}$.
This can be considered the scaling of the memory size, because we need this amount to store all the variables.
Note that this memory scaling is much better than $2^{N_{\rm bos}n_{\rm q}}$, which is required for the direct computation of the eigenvalues with exact diagonalization methods.
For practical applications, $n_{\rm q}$ does not increase with $N_{\rm bos}$ (each bosonic degree of freedom is represented with the same number of qubits), and hence the necessary memory size increase only linearly with the system size.
In addition to the memory scaling, the computational cost for one sweep scales polynomially with $N_{\rm bos}$ and $a_{\rm dig}$.
 
Strictly speaking, it is difficult to establish the exponential suppression of the truncation effect unless a clear exponential behavior sets in at a small truncation level because exponentially large statistics and hence exponentially large computational costs are needed to have exponentially small statistical error.\footnote{In the Monte Carlo simulation, the statistical error scales as $({\rm number\ of\ configurations})^{-1/2}$.}
However, to establish that the error decays faster than a certain power, only a polynomially large cost is needed.

\subsection{Single boson \texorpdfstring{($N_{\rm bos}=1$)}{Nbos=1}}\label{sec:single-boson-simulation}
\hspace{0.51cm}
As a sanity check of the method, let us consider the case of a single boson. 
In this case, the exact diagonalization method can be used to determine the low-energy spectrum rather accurately even for large truncation level $\Lambda$. Therefore, we can confirm that the values obtained from the Monte Carlo simulations described in the previous sections are correct. 

We considered the Hamiltonian with the following quartic interaction:
\begin{align}\label{eq:quartic-potential}
    V(\hat{x}) = \frac{\lambda}{4}\hat{x}^4+\frac{m^2}{2}\hat{x}^2 \, .
\end{align}

Note that we can consider negative $m^2$ as well. 
We studied $\lambda=1.0$, $m^2=\pm 1.0$ and $a_{\rm dig}=0.3, 0.5, 0.7$. 
The step size $\Delta$ is set to $0.001$ so that $\frac{\Delta}{a_{\rm dig}^2}$ is small and the error associated with the truncation of the Taylor expansion of $e^{-\Delta\cdot\frac{\hat{p}^2}{2}}$ is suppressed.
To focus on the errors coming from $a_{\rm dig}$ and not from $R$, we took $R$ and $\Lambda$ to be very large (specifically, $\Lambda=2001$ and $R=1000a_{\rm dig}$).

\begin{figure}[htbp]
  \begin{center}
    \includegraphics{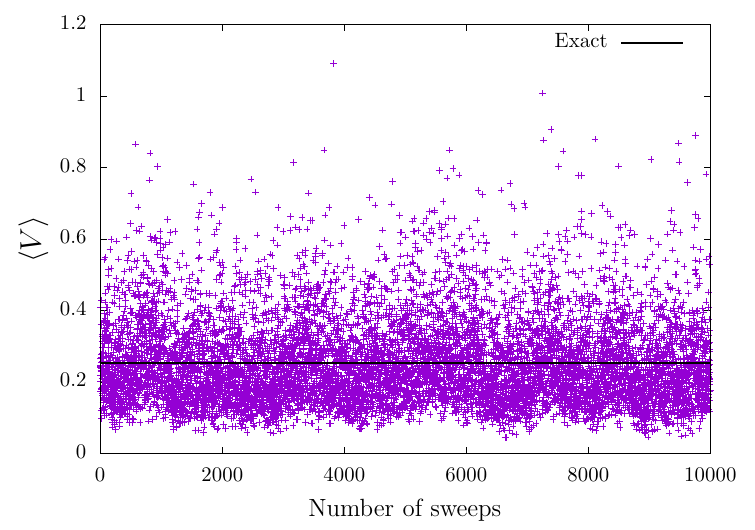}
  \end{center}
  \caption{
  Simulation history of $T=0.1$, $a_{\rm dig}=0.5$, $\Delta=0.001$, $B_{\rm max}=\frac{K}{2}=5000$. 
  The exact value obtained by exact diagonalization is 0.2539 in this case.
  }\label{fig:V-for-quartic-potential_history}
\end{figure}

\begin{figure}[htbp]
  \begin{center}
   \includegraphics{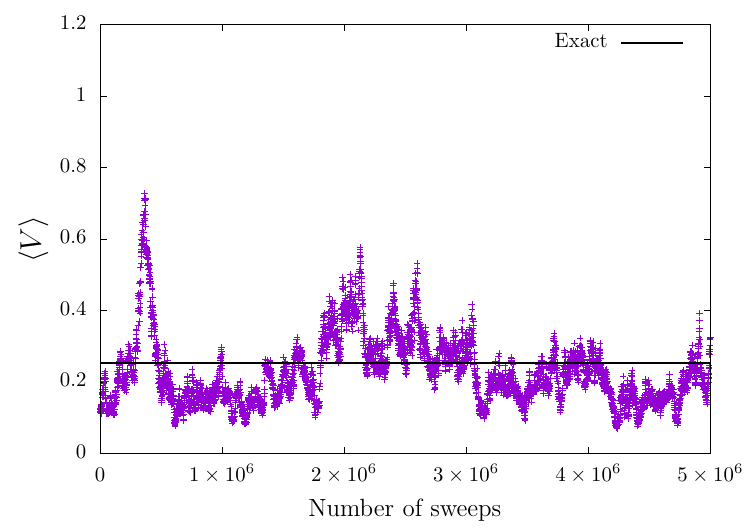}
  \end{center}
  \caption{Simulation history of $T=0.1$, $a_{\rm dig}=0.5$, $\Delta=0.001$, $B_{\rm max}=1$.
  Autocorrelation is much longer than the simulation with $B_{\rm max}=\frac{K}{2}=5000$ shown in Fig.~\ref{fig:V-for-quartic-potential_history}.
  }\label{fig:V-for-quartic-potential_history_non_cluster}
\end{figure}

The results are summarized in Table~\ref{table:V-for-quartic-potential}. 
`Exact' values are obtained by exactly diagonalizing the corresponding Hamiltonian in the truncated coordinate basis, with a sufficiently large value of $\Lambda$ such that the first four nonzero digits are obtained precisely. 
Both methods give consistent results by taking into account the stochastic (statistical) error from the MCMC simulation. 
The simulation history for $a_{\rm dig}=0.5$, $m^2=+1.0$ is shown in Fig.~\ref{fig:V-for-quartic-potential_history}. 
For this plot, we set the maximum cluster size that is updated simultaneously to be $B_{\rm max}=\frac{K}{2}$. 
The advantage of updating a large cluster simultaneously can be understood by comparing this plot with Fig.~\ref{fig:V-for-quartic-potential_history_non_cluster} corresponding to a simulation with $B_{\rm max}=1$. 

\begin{table}
\begin{center}
\begin{tabular}{ |c||c|c||c|c| }
\hline
$a_{\rm dig}$ & $m^2=1.0$, MC & $m^2=1.0$, Exact & $m^2=-1.0$, MC & $m^2=-1.0$, Exact\\ 
\hline
\hline
$0.3$ & $0.2626(26)$ & $0.2618$ & $-0.06326(68)$ & $-0.06354$ \\
\hline
$0.5$ & $0.2533(53)$ & $0.2539$ &  $-0.0664(27)$ & $-0.06633$ \\  
\hline
$0.7$ &  $0.2482(70)$ & $0.2414$ & $-0.0717(18)$  & $-0.07024$\\  
\hline
\end{tabular}
 \end{center}
\caption{
 $\langle V \rangle_{T=0.1}$ from Monte Carlo (MC) and exact diagonalization (Exact) for 
$V(x)=\frac{1}{4}x^4+\frac{m^2}{2}x^2$, $m^2=\pm 1.0$, $T=0.1$, various choices of $a_{\rm dig}$, with the infrared cutoff $R=1000a_{\rm dig}$ ($\Lambda=2001$). 
 For MC, $\Delta=0.001$ was used and $10^4$ sweeps were performed.
 }\label{table:V-for-quartic-potential}
\end{table}

\subsection{Scalar QFT \texorpdfstring{($N_{\rm bos}=L^d$)}{Nbos=Ld}}\label{sec:2d-lattice-simulation}
\hspace{0.51cm}
Let us consider the scalar QFT on a square $d$-dimensional lattice of $L$ sites in each direction. 
The number of bosons is $N_{\rm bos}=L^d$, which quickly grows and makes it impossible to construct the Hilbert space explicitly on classical devices. 

To demonstrate the validity of the method, we consider the free theory with the potential:
\begin{align}\label{eq:free-field-potential}
    V = \frac{m_{\rm lat}^2}{2}\sum_{\vec{n}}\hat{\phi}_{\vec{n}}^2\, , 
\end{align}
where $m_{\rm lat}=a_{\rm lat}m$ in the Hamiltonian defined by \eqref{eq:2d-lattice-Hamiltonian}. We focus on quantities that can be analytically computed for the case of the infinite-dimensional local Hilbert space (the full QFT without digitization effects) and we reproduce those results by taking the limit of $a_{\rm dig}\to 0$ in our Monte Carlo simulations.
(We use this particular setup because the analysis is simpler and the essence of the MCMC approach can be conveyed efficiently. 
We will comment on the cases without analytic results later in this section.)

The Fourier transform on a lattice is defined by 
\begin{align}\label{eq:fourier-transform}
    &
    \hat{\tilde{\phi}}_{\vec{q}}
    =
    \frac{1}{\sqrt{L^{d}}}
    \sum_{\vec{n}}
    e^{-i\vec{q}\cdot\vec{n}}\hat{\phi}_{\vec{n}},
    \qquad
    \hat{\phi}_{\vec{n}}
    =
    \frac{1}{\sqrt{L^{d}}}
    \sum_{\vec{q}}
    e^{i\vec{q}\cdot\vec{n}}\hat{\tilde{\phi}}_{\vec{q}}\ , 
    \nonumber\\
    &
    \hat{\tilde{\pi}}_{\vec{q}}
    =
    \frac{1}{\sqrt{L^{d}}}
    \sum_{\vec{n}}
    e^{-i\vec{q}\cdot\vec{n}}\hat{\pi}_{\vec{n}},
    \qquad
    \hat{\pi}_{\vec{n}}
    =
    \frac{1}{\sqrt{L^{d}}}
    \sum_{\vec{q}}
    e^{i\vec{q}\cdot\vec{n}}\hat{\tilde{\pi}}_{\vec{q}} \, , 
\end{align}
where $\vec{q}=(q_1,\cdots,q_d)$, $q_j=\frac{2\pi}{L}\ell_j$, $\ell_j=1,2,\cdots,L$. 
We can write the free Hamiltonian in terms of $\hat{\tilde{\phi}}_{\vec{q}}$ and $\hat{\tilde{\pi}}_{\vec{q}}$ as
\begin{align}\label{eq:free-field-hamiltonian}
    \hat{H}_{\textrm{lat, free}}
    &=
    \sum_{\vec{n}}\left(
    \frac{1}{2}
    \hat{\pi}_{\vec{n}}^2
    +
    \frac{1}{2}
    \sum_{\mu=1}^d
    \left(
    \hat{\phi}_{\vec{n}+\hat{\mu}}-\hat{\phi}_{\vec{n}}
    \right)^2
    +
    \frac{m_{\rm lat}^2}{2}\hat{\phi}_{\vec{n}}^2
    \right)
    \nonumber\\
    &=
    \sum_{\vec{q}}\left(
    \frac{1}{2}
    \hat{\tilde{\pi}}_{\vec{q}}\hat{\tilde{\pi}}_{-\vec{q}}
    +
    \left(
    2
    \sum_{\mu=1}^d\sin^2\left(\frac{q_\mu}{2}\right)
    +
    \frac{m_{\rm lat}^2}{2}
    \right)
    \hat{\tilde{\phi}}_{\vec{q}}\hat{\tilde{\phi}}_{-\vec{q}}
    \right)
    \nonumber\\
    &=
    \sum_{\vec{q}}\left(
    \frac{1}{2}
    \hat{\tilde{\pi}}_{\vec{q}}\hat{\tilde{\pi}}_{-\vec{q}}
    +
    \frac{\omega^2_{{\rm lat},\vec{q}}}{2}
    \hat{\tilde{\phi}}_{\vec{q}}\hat{\tilde{\phi}}_{-\vec{q}}
    \right)\ , 
\end{align}
where
\begin{align}\label{eq:frequency-modes}
    \omega^2_{{\rm lat},\vec{q}}
    =
    m_{\rm lat}^2
    +
    4
    \sum_{\mu=1}^d\sin^2\left(\frac{q_\mu}{2}\right) \, . 
\end{align}

Each mode contributes $\frac{\omega_{\vec{q}}}{2}$ to the ground-state energy.
The zero-point fluctuation of each mode is 
\begin{align}\label{eq:zero-mode}
    \left\langle
    \hat{\tilde{\phi}}_{\vec{q}}\hat{\tilde{\phi}}_{-\vec{q}}
    \right\rangle
    =
    \frac{1}{2\omega_{{\rm lat},\vec{q}}} \, . 
\end{align}
This exact value should be obtained in the limits $\Delta\to 0$, $R\to\infty$, $a_{\rm dig}\to 0$ and $T\to 0$.
Note that we need to take both the digitization spacing to zero, and the energy scale to zero.
On the other hand, at finite temperatures we have:
\begin{align}\label{eq:width-mom-rep}
    \left\langle
    \hat{\tilde{\phi}}_{\vec{q}}\hat{\tilde{\phi}}_{-\vec{q}}
    \right\rangle
    =
    \frac{e^{\beta\omega_{\vec{q}}/2}+e^{-\beta\omega_{\vec{q}}/2}}{e^{\beta\omega_{\vec{q}}/2}-e^{-\beta\omega_{\vec{q}}/2}}\times\frac{1}{2\omega_{{\rm lat},\vec{q}}}
    =
    \frac{1}{2\omega_{{\rm lat},\vec{q}}\tanh(\beta\omega_{\vec{q}}/2)} \, .
\end{align}

To reproduce this result, we only take the limits $\Delta\to 0$, $R\to\infty$ and $a_{\rm dig}\to 0$.
We use this relation for the numerical demonstrations that follow.
Note that the digitization and Fourier transform do not commute in general.
Therefore, even for the free theory, the estimation of the digitization effect is not trivial.  

The ground-state wave function for each mode $\tilde{\phi}_{\vec{q}}$ is Gaussian with the width $\frac{1}{\sqrt{\omega_{{\rm lat}, \vec{q}}}}$. 
For the higher-momentum modes, the widths are narrower, and smaller $a_{\rm dig}$ will be needed for better approximation. 

For numerical demonstration, we study a two-dimensional $4\times 4$ lattice. 
The number of bosons is $N_{\rm bos}=4\times 4=16$ and the dimension of the Hilbert space $\Lambda^{N_{\rm bos}}$ increases so quickly with $\Lambda$ that numerical analysis on classical devices is practically impossible if the Hilbert space is constructed explicitly.

We choose the parameters to be $a_{\rm lat}=1$ and $m^2=1$. 
For $\vec{q}=(q_x,q_y)=(0,0)$ and $(\pi,\pi)$, $\frac{1}{2\omega_{{\rm lat}, \vec{q}}}$ is $\frac{1}{2}$ and $\frac{1}{6}$, respectively. 
Combined with \eqref{eq:width-mom-rep}, the values shift to $\frac{1}{2\tanh(0.5)}\simeq1.0819$ and $\frac{1}{6\tanh(1.5)}\simeq 0.1841$, respectively, at $T=1$.
We estimate the truncation effects on these values. 

\begin{figure}[htbp]
  \begin{center}
  \includegraphics{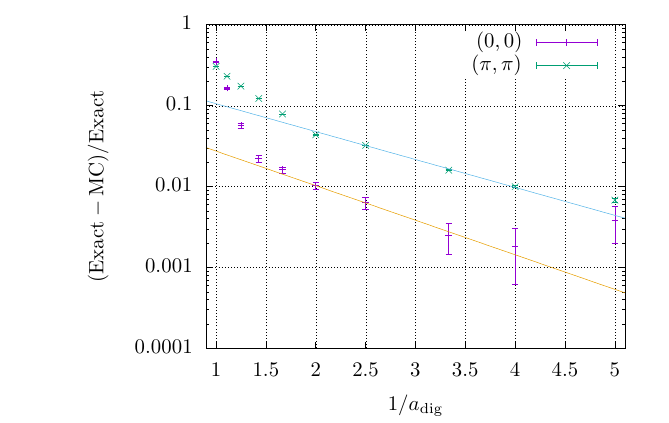}
  \end{center}
  \caption{
  By using the values in Table~\ref{table:a-dependence-2d-lattice-4*4}, 
  $\frac{{\rm Exact}-{\rm MC}}{{\rm Exact}}$ is plotted, where `${\rm Exact}$' is the exact value \eqref{eq:width-mom-rep} which should be obtained in the limit of $\Delta\to 0$, $R\to\infty$ and $a_{\rm dig}\to 0$ and `${\rm MC}$' is the numerical results in Table~\ref{table:a-dependence-2d-lattice-4*4}.
  The horizontal axis is $\frac{1}{a_{\rm dig}}$. 
  We performed a fit of the data in Table~\ref{table:a-dependence-2d-lattice-4*4} using the function $\left\langle\hat{\tilde{\phi}}_{\vec{q}}\hat{\tilde{\phi}}_{-\vec{q}}\right\rangle=Ae^{-B/a_{\rm dig}}+C$. In this plot we fix $C$ to the exact value $1.081977$ for $\vec{q}=(0,0)$ and $0.184131$ for $\vec{q}=(\pi,\pi)$. This is different from Fig.~\ref{fig:2d-scalar-log-2}, in which $C$ is treated as a fitting parameter.
  We obtained $A=-0.079(11)$, $B=0.983(66)$ for $\vec{q}=(0,0)$ from $a_{\rm dig}=0.25, 0.30, 0.40, 0.50$ and $A= -0.0430(69)$, $B= 0.794(58)$ for $\vec{q}=(\pi,\pi)$ from $a_{\rm dig}=0.20, 0.25, 0.30, 0.40$. 
}\label{fig:2d-scalar-log}
\end{figure}

We took $\frac{\Delta}{2a_{\rm dig}^2}$ to be $0.01$ at maximum. Therefore, we expect the error associated with Trotterization is at most a few percent. 
The values we obtained are shown in Table~\ref{table:a-dependence-2d-lattice-4*4}. 
In Fig.~\ref{fig:2d-scalar-log}, we plot $\frac{{\rm Exact}-{\rm MC}}{\rm Exact}$, where `${\rm Exact}$' is the exact value \eqref{eq:width-mom-rep} which should be obtained in the limit of $\Delta\to 0$, $R\to\infty$ and $a_{\rm dig}\to 0$, and `${\rm MC}$' is the numerical result in Table~\ref{table:a-dependence-2d-lattice-4*4}.
The horizontal axis is $\frac{1}{a_{\rm dig}}$. 
We can see that the error decreases to $\frac{{\rm Exact}-{\rm MC}}{{\rm Exact}}\sim 0.01$, which is more or less the expected value of the Trotterization effect.  

Very roughly, the width of the distribution of $\tilde{\phi}_{\vec{q}}$ is estimated by $\sqrt{\langle\hat{\tilde{\phi}}_{\vec{q}}\hat{\tilde{\phi}}_{-\vec{q}}\rangle}$, which is $\sqrt{1.0819}\simeq 1.04$ for $\vec{q}=(q_x,q_y)=(0,0)$ and $\sqrt{0.1841}\simeq 0.43$ for $\vec{q}=(q_x,q_y)=(\pi,\pi)$. A natural expectation is that the error associated with the digitization disappears exponentially fast once $a_{\rm dig}$ becomes smaller than these values. Qualitatively, we can see such an exponential decrease in Fig.~\ref{fig:2d-scalar-log}.

\begin{table}
\begin{center}
\begin{tabular}{ |c|c|c||c|c||c|c|c|c| }
\hline
$a_{\rm dig}$ & $\Delta$ & $K=\beta/\Delta$ &$\vec{q}=(0,0)$ & $\vec{q}=(\pi,\pi)$ 
& $d_{(0,0)}$ & $d_{(\pi,\pi)}$ & $n_\mathrm{stream}$ & $n_\mathrm{step}$\\ 
\hline
\hline
0.20 & 0.0008 & 1250 & 1.0778(20) & 0.18288(8) & 37 & 4 & 40 & $3.93\times10^5$ \\
\hline
0.25 & 0.00125 & 800 & 1.0800(13) & 0.18230(5) & 38 & 4 & 39 & $1.1\times10^6$ \\
\hline
0.30 & 0.0015 & 667 & 1.0793(11) & 0.18120(4) & 37 & 4 & 22 & $2.57\times10^6$ \\
\hline
0.40 & 0.002 & 500 & 1.0751(12) & 0.17819(4) & 41 & 4 & 19 & $2.59\times10^6$ \\
\hline
0.50 & 0.002 & 500 & 1.0709(10) & 0.17611(3) & 59 & 5 & 10 & $1\times10^7$ \\
\hline
0.60 & 0.005 & 200 & 1.0645(15) & 0.16961(4) & 130 & 7 & 10 & $1\times10^7$ \\
\hline
0.70 & 0.005 & 200 & 1.0579(26) & 0.16156(6) & 369 & 15 & 10 & $1\times10^7$ \\
\hline
0.80 & 0.005 & 200 & 1.0208(39) & 0.15214(10) & 968 & 55 & 10 & $1.1\times10^7$ \\
\hline
0.90 & 0.005 & 200 & 0.9039(55) & 0.14167(17) & 2174 & 148 & 10 & $1.1\times10^7$ \\
\hline
1.00 & 0.005 & 200 & 0.7038(70) & 0.12792(25) & 4491 & 319 & 10 & $1\times10^7$ \\
\hline
\end{tabular}
 \end{center}
\caption{
  $\left\langle
\hat{\tilde{\phi}}_{\vec{q}}\hat{\tilde{\phi}}_{-\vec{q}}
\right\rangle$ for $\vec{q}=(q_x,q_y)=(0,0)$ and $(\pi,\pi)$ on 2d $4\times 4$ lattice, $a_{\rm lat}=1$, $m^2=1$, $\lambda=0$, and $T=1$ via MCMC. 
$R=a_{\rm dig}\times 1000$, $\Lambda=2001$. We chose $\Delta$ such that $\frac{\Delta}{2a_{\rm dig}^2}\le 0.01$. 
The exact value without digitization ($\Delta\to 0$ followed by $a_{\rm dig}\to 0$) is $\frac{1}{2\tanh(0.5)}\simeq1.081977$ and $\frac{1}{6\tanh(1.5)}\simeq 0.184131$. 
Simulations are conducted by using multiple ($n_\mathrm{stream}$) streams with different random seeds. Each stream has $n_\mathrm{step}\geq 3.93\times 10^5$ steps.
The auto-correlation length for each $\vec{q}$ is estimated by computing the integrated auto-correlation time for each stream, and for the largest length $d_{\vec{q}}$, the initial $10d_{\vec{q}}$ steps are discarded as a burn-in period, and the following steps are split into $10d_{\vec{q}}$-step samples.
$d_{\vec{q}}<10^3$ for $a\leq 0.8$, and for $a=0.9$ and $1.0$, stream sizes are greater than $2200d_{\vec{q}}$.
The number shown in the parenthesis indicates the unbiased estimation of the standard deviation from the averages of at least ten streams.}
\label{table:a-dependence-2d-lattice-4*4}
\end{table}

\begin{figure}[htbp]
  \begin{center}
  \includegraphics{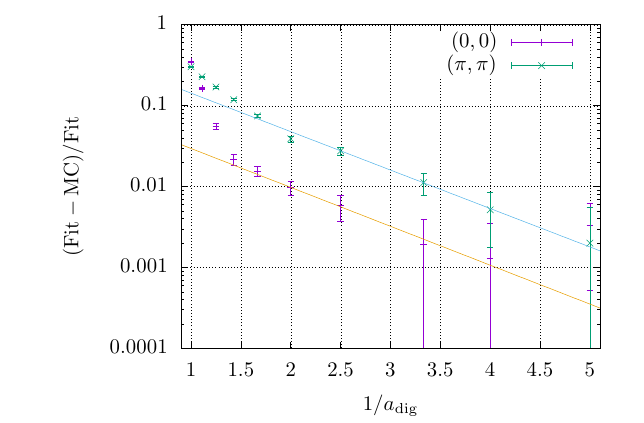}
  \end{center}
  \caption{
  By using the values in Table~\ref{table:a-dependence-2d-lattice-4*4}, we performed a fit using the function $\left\langle\hat{\tilde{\phi}}_{\vec{q}}\hat{\tilde{\phi}}_{-\vec{q}}\right\rangle=Ae^{-B/a_{\rm dig}}+C$. 
  Unlike Fig.~\ref{fig:2d-scalar-log}, here we treated $C$ as a fitting parameter.
  For $\vec{q}=(0,0)$, we obtained $A=-0.096(47)$,  $B=1.10(28)$ and  $C=1.0814(11)$ from $a_{\rm dig}=0.25, 0.30, 0.40, 0.50$.
  For $\vec{q}=(\pi,\pi)$, we obtained $A=-0.0780(85)$, $B=1.094(50)$ and  $C=0.18324(10)$ from $a_{\rm dig}=0.20, 0.25, 0.30, 0.40$.
  We plot $\frac{{\rm Fit}-{\rm MC}}{{\rm Fit}}$, where `${\rm Fit}$' is the fit value $C$.
  This is again in contrast to Fig.~\ref{fig:2d-scalar-log} where we used the `${\rm Exact}$' value for $C$.
  The horizontal axis is $\frac{1}{a_{\rm dig}}$. 
}\label{fig:2d-scalar-log-2}
\end{figure}

For the example discussed above, we knew the analytic results in the limit of $a_{\rm dig}\to 0$.
Even without knowing the analytic result, the digitization error analysis is straightforward, because our method works for any $R$ and $a_{\rm dig}$. 
For simplicity, let us assume that $R=\infty$ and focus on the finite-$a_{\rm dig}$ effect, as above. 
For various values of $a_{\rm dig}$, we can calculate $\langle\mathcal{O}\rangle$, where $\mathcal{O}$ is observable under consideration, say $\phi^2$. 
Then, we can determine the $a_{\rm dig}$-dependence by fitting the numerical results at finite $a_{\rm dig}$ values. 
If the analytic result at $a_{\rm dig}=0$ is known, such a fit is easier (has less free parameters), but the fit can be conducted even without knowing the value at $a_{\rm dig}=0$.
In Fig.~\ref{fig:2d-scalar-log-2} we show how to perform a fit that includes the $a_{\rm dig}=0$ value of the observable as a free parameter.
We compare the fitted result with the known $a_{\rm dig}=0$ result and find that they are statistically compatible.
Alternatively, even if an analytical result is not available, one can determine the value at $a_{\rm dig}=0$ (without any digitization of the degrees of freedom) by performing the standard Euclidean path integral computation via MCMC.

\section{Conclusion and discussion}\label{sec:conclusion}
\hspace{0.51cm}
In this paper, we introduced an MCMC-based technique to determine the digitization effects in a class of bosonic systems, which have the Hamiltonian of the form \eqref{eq:general-bosonic-Hamiltonian}, in the coordinate-basis truncation scheme. 
This technique enables us to study the expectation values of various operators at finite temperature including their digitization effects.
By dialing the temperature, various energy scales can be studied.
As a prototypical QFT application, we studied the $(2+1)$-dimensional scalar QFT regularized on a lattice.
To check its convergence to the right answer, we studied the weak-coupling limit.
The inclusion of the interactions is straightforward, as we have demonstrated in the case of a single-boson system.
Our method can be used to estimate the resources needed for realistic quantum simulations, and also, to check the validity of the results of the quantum simulations.   
As a specific example relevant for the NISQ era, we can consider a variational algorithms to determine low-energy quantum states of a multi-dimensional lattice system.
By using the MCMC method, one determines the ground-state energy of the digitized theory, including the scaling of the digitization effects.
Whether this value can be reproduced is a good test of variational algorithms on NISQ devices when no other classical methods can be used due to the exponentially increasing requirements with the lattice size.
Once we could find a quantum algorithm and quantum device that can pass this benchmark test, then we can use it to study more complicated observables which cannot be accessed by the MCMC method, such as excited-state energies or time-dependent correlation functions.

In this paper, we took the `infrared'-cutoff parameter $R$ very large and focused on the effects due to finite $a_{\rm dig}$.
In near-term quantum simulations the finite-$R$ effects can also be relevant.
It is straightforward to study small values of $R$ and estimate the finite-$R$ effects, by using the same simulation technique we have proposed.

To apply the method introduced in this paper, each bosonic variable must take values in $\mathbb{R}$.
An interesting class of theories of this kind is Hermitian multi-matrix models, which are important in the context of the holographic duality.
A nontrivial issue for those models is how to reconcile the digitization of the gauge field with the existence of the gauge symmetry.
However, the study of the corresponding ungauged models~\cite{Maldacena:2018vsr, Berkowitz:2018qhn, Pateloudis:2022oos}, which are also relevant for the duality, is straightforward.
It is not necessarily a problematic issue if one considers a quantum simulation on the extended Hilbert space, which is a rather common approach.
For such an approach to work, one has to make sure that the low-energy states of the ungauged model in the extended Hilbert space are correctly described.
Lattice gauge theories form another interesting class. 
Although unitary variables which appear in the Kogut-Susskind formulation~\cite{Kogut:1974ag} of lattice gauge theory cannot be studied by using the formulation given in this paper, it is straightforward to study the orbifold lattice~\cite{Kaplan:2002wv,Buser:2020cvn}.

Whether it is possible to study other digitization schemes by Monte Carlo methods is an interesting question. For SU($N$) gauge theories, digitization schemes based on discrete subgroups have been proposed~\cite{Alexandru:2019nsa,Ji:2020kjk,Ji:2022qvr,Haase:2020kaj}.
The effects of such schemes can be studied by using the standard lattice simulations~\cite{Hackett:2018cel,Alexandru:2021jpm} and in principle, it is possible to take the continuum limit along the time directions both for gauged and ungauged theories.
While such digitization schemes lack systematic un-digitization limits, they might be sufficiently good tools in the NISQ era.
Another interesting approach that can be studied by Monte Carlo is the use of non-commutative geometry. 
For example, Refs.~\cite{Alexandru:2019ozf,Alexandru:2022son} studied the truncation of the target space of the O(3) nonlinear sigma model by fuzzy sphere. 

Another potentially useful direction is the use of variational Monte Carlo methods, specifically with the neural quantum states.
Such an approach, which was confirmed to be valid for some theories before digitization~\cite{Han:2019wue,Rinaldi:2021jbg,Stokes:2021wnv,Pescia:2021kxb}, may work even with digitization, and the introduction of fermions is straightforward at least conceptually. 
The introduction of the singlet constraint is also straightforward, for example by adding the square of the generators of gauge symmetry to the action as a penalty term. 
The MCMC technique introduced in this paper could be used for cross checking purposes.

\begin{center}
\textbf{Acknowledgements}
\end{center}
\hspace{0.51cm}
The authors would like to thank H.~Lamm for useful comments. 
M.~H. and E.~R. thank the Royal Society International Exchanges award IEC/R3/213026. 
M.~H. and J.~L. acknowledge support from qBraid Co.
M.~H. thanks the STFC Ernest Rutherford Grant ST/R003599/1.
E.~R. was in part supported by Nippon Telegraph and Telephone Corporation (NTT) Research during the initial stages of this project.
J.~L. is supported in part by International Business Machines (IBM) Quantum through the Chicago Quantum Exchange, and the Pritzker School of Molecular Engineering at the University of Chicago through AFOSR MURI (FA9550-21-1-0209).
M.~T. was partially supported by the Japan Society for the Promotion of Science (JSPS) Grants-in-Aid for Scientific Research (KAKENHI) Grants No. JP20K03787 and JP21H05185.

\begin{center}
\textbf{Data management}
\end{center}
\hspace{0.51cm}
No additional research data beyond the data presented and cited in this work are needed to validate the research findings in this work. 
Simulation data are openly available at the following URL/DOI:
\url{https://zenodo.org/record/7766237}.

\appendix
\section{Short review of MCMC}\label{sec:MCMC}
Markov Chain Monte Carlo (MCMC) is a class of theories that are used to create a probability distribution efficiently. 
Specifically, let us consider $n$ variables $x_1,\cdots,x_n$ which may be real numbers or integers. 
The goal is to obtain a sequence of `configurations' $\{x^{(i)}\}=(x_1^{(i)},\cdots,x_n^{(i)})$ ($i=1,2,\cdots$) whose distribution converges to the target probability distribution $P(x_1,\cdots,x_n)$. 
Such a sequence allows us to calculate the expectation value of a function $f(x_1,\cdots,x_n)$ as 
\begin{align}\label{def:mc-expectation-value}
    \langle f(x_1,\cdots,x_n)\rangle
    &\equiv
    \int dx_1\cdots dx_n
    f(x_1,\cdots,x_n)P(x_1,\cdots,x_n)
    \nonumber\\
    &=
    \lim_{N_{\rm config}\to\infty}
    \frac{1}{N_{\rm config}}\sum_{i=1}^{N_{\rm config}}
    f(x_1^{(i)},\cdots,x_n^{(i)}). 
\end{align}

The MCMC algorithms~\cite{Metropolis:1953am,rosenbluth1953proof} are designed so that the sequence is a Markov chain, i.e. the probability of obtaining $\{x^{(i+1)}\}$ after $\{x^{(i)}\}$ depends only on $\{x^{(i)}\}$ and does not depend on $\{x^{(i-1)}\},\{x^{(i-2)}\},\cdots$, and the transition probability between configurations denoted by $T(\{x\}\to\{x'\})$ satisfies the following three conditions:
\begin{enumerate}
\item
\textbf{Irreducibility}. The Markov chain defined by $T(\{x\}\to\{x'\})$ is irreducible, i.e., transition between any pair $\{x\}$ and $\{x'\}$ is possible with a finite number of steps. 

\item
\textbf{Aperiodicity}.
The greatest common divisor of the numbers of steps needed for a transition from $\{x\}$ to itself is called the period. 
The transition probability $T(\{x\}\to\{x'\})$ is chosen in such a way that the Markov chain defined is aperiodic, i.e,  the period is $1$ for any $\{x\}$.

\item\
\textbf{Detailed balance condition} is satisfied, i.e., 
\begin{align}\label{def:detailed-balance}
    P(\{x\})\cdot T(\{x\}\to\{x'\})
    =
    P(\{x'\})\cdot T(\{x'\}\to\{x\}). 
\end{align}
for any $\{x\}$ and $\{x'\}$. 
\end{enumerate}

In the Metropolis algorithm~\cite{Metropolis:1953am,rosenbluth1953proof}, transition from $\{x\}$ to $\{x'\}$ and that from $\{x'\}$ to $\{x\}$ are proposed with the same probability, and the proposal $\{x\}$ to $\{x'\}$ is accepted with the probability ${\rm min}(1,P(\{x'\})/P(\{x\}))$. 
It is straightforward to check that the detailed balance condition is satisfied. 
Whether the other two conditions are satisfied depends on the details of the transition probability $T(\{x\}\to\{x'\})$. 

In the MCMC algorithms, most of the computational resources are used to create \textit{important} configurations dominating the probability distribution. This feature, which is called importance sampling, allows us to drastically save simulation time.

\providecommand{\href}[2]{#2}\begingroup\raggedright\endgroup

\end{document}